\documentclass[aps,pra,showpacs,twocolumn,superscriptaddress]{revtex4}

\usepackage{graphicx,mathptmx}
\usepackage{epsfig}

\usepackage{amsmath}
\usepackage{amssymb}

\usepackage[dvipdfm]{hyperref}
\hypersetup{colorlinks=true,linkcolor=blue,citecolor=blue,urlcolor=blue}

\begin{document}


\title{Bell tests with random measurements require very high detection efficiencies}


\date{\today}


\author{Esteban~S.~G\'{o}mez}
\author{Gustavo~Ca\~{n}as}
\author{Johanna~F.~Barra}
\affiliation{Center for Optics and Photonics, MSI-Nucleus on Advanced Optics, Departamento de F\'{\i}sica, Universidad de Concepci\'{o}n, 160-C Concepci\'{o}n, Chile}
\author{Ad\'{a}n~Cabello}
\affiliation{Departamento de F\'{\i}sica Aplicada II, Universidad de Sevilla, E-41012 Sevilla, Spain}
\affiliation{Universidade Federal de Minas Gerais,
 Caixa Postal 702, 30123-970, Belo Horizonte, MG, Brazil}
\author{Gustavo~Lima}
\affiliation{Center for Optics and Photonics, MSI-Nucleus on Advanced Optics, Departamento de F\'{\i}sica, Universidad de Concepci\'{o}n, 160-C Concepci\'{o}n, Chile}



\begin{abstract}
The observation that violating Bell inequalities with high probability is possible even when the local measurements are randomly chosen,
as occurs when local measurements cannot be suitably calibrated or the parties do not share a common reference frame, has recently attracted much theoretical and experimental efforts. Here we show that this observation is only valid when the overall detection efficiency is very high ($\eta \geq 0.90$), otherwise, even when using the highest detection efficiency of recent photonic Bell tests, the probability of demonstrating nonlocality is negligible (e.g., it is smaller than $0.02\%$ for $\eta=0.785$). Our results show that detection efficiency is a much more critical resource for real-world applications than it was previously thought.
\end{abstract}

\pacs{03.65.Ud,42.50.Xa}



\maketitle


\section{Introduction}

Bell inequalities define constraints on the probabilities of events in experiments involving local measurements on composite systems. These constraints are satisfied by local hidden variable (LHV) theories and, under the appropriate conditions, are violated by quantum mechanics \cite{Bell}. Experimental violations of Bell inequalities (i.e., Bell tests) are fundamental for revealing quantum nonlocality \cite{Bell}, certifying secure communications \cite{Ekert91}, better-than-classical distributed computation \cite{BCMD10}, randomness \cite{Acin}, and entanglement \cite{GT09}.

The realization of a loophole-free Bell test (i.e., a Bell test without requiring extra assumptions) is probably the most important experimental problem in fundamental quantum mechanics. Despite an enormous progress in recent years \cite{Lima2010,Chaves11,Smith2012,Zeilinger12,CS12,GMRWKLCGNUZ12}, such a test is still missing. The reason is that there are many challenges associated with loophole-free violations of Bell inequalities. First, one observer's local measurement result must be outside the light cone of the other observer's measurement choice. The idea is to prevent influences at the speed of light between these events. Thus, they must be spatially separated. In this case, the experiment is said to be free of the locality loophole \cite{Aspect,Weihs}. A second requirement is that in a Bell test it is essential to guarantee that the overall detection efficiency $\eta$ (defined as the ratio between the detected and emitted particles) is above a threshold value, hard to achieve experimentally with photons. Otherwise, one can explain with LHV theories why the detected subensemble (apparently) violates the Bell inequality \cite{Pearle}.

Another experimental challenge is that Bell tests require that the local observers implement well-calibrated measurements and share a common reference frame; otherwise, the violation decreases and may eventually vanish. The observation that calibrated local devices and a common reference frame count as resources stimulated the research on quantum nonlocality free of a common reference frame \cite{Adan03,Cabello03,Bartlett10,Bartlett11,Bartlett12,Brunner11,Brunner12}. The most recent approach to the problem is based on a simple observation: when one considers a standard Bell test but uses randomly chosen local measurements, then the probability $P_{\rm viol}$ that the correlations between the results violate a Bell inequality can be very high \cite{Bartlett10}. Due to the fundamental importance of Bell tests and their applications, this observation has motivated new types of Bell experiments \cite{Brunner11,Wallman}.

However, so far, this approach has not taken into account the problem of the detection efficiency. The question is whether the conclusions of these theoretical and experimental works hold when realistic detection efficiencies are considered. Here we study how the detection efficiency affects the conclusions reached for the Bell tests of the Clauser-Horne-Shimony-Holt (CHSH) inequality \cite{CHSH,ClauserHorne} with random local measurements \cite{Bartlett10,Bartlett11,Bartlett12,Brunner11,Brunner12,Wallman}. We focus on this Bell inequality because it is the simplest one, the most frequently used for real-world applications, and its violation requires detection efficiencies smaller than almost any other bipartite Bell inequality \cite{BrunnerPLA,CRV08,Vertesi2009,CLR09}.

We study three different types of Bell tests with random measurements. First, the two types defined in \cite{Bartlett10}. In the first scenario, one considers random isotropic measurements (RIM) in which each party chooses both measurement directions randomly, independently, and uniformly distributed over the Bloch sphere. In the second scenario, one considers random orthogonal measurements (ROM) in which one of the measurement directions of each party is randomly, independently, and uniformly distributed over the Bloch sphere, while the second measurement is also random but orthogonal to the first one. The interest in the ROM scenario is that it allows for a higher probability of violating the CHSH inequality. Notice that in both scenarios the parties do not share a reference frame. However, the ROM scenario assumes that devices are perfectly calibrated. For the CHSH inequality, and {\em assuming perfect detection efficiency} (i.e., $\eta=1$), $P_{\rm viol}=0.28$ for the RIM and $P_{\rm viol}=0.41$ for the ROM scenario of random Bell tests \cite{Bartlett10}.

We also study a third type of random Bell tests proposed in \cite{Bartlett12,Brunner11}, which is a variation of the ROM scenario where instead of performing two orthogonal measurements, the parties use a random orthogonal triad measurements (ROTM). In this scenario Alice and Bob can perform three orthogonal measurements. For the CHSH inequality only two measurement settings are required, but due to the extra number of measurements, there are more equivalent forms for the inequality. This increases the probability of violating the inequality and the interesting point of the ROTM scenario is that it allows $P_{\rm viol}=1$ when $\eta=1$.


\section{Method}

To test the CHSH inequality, each party (Alice and Bob) performs two-outcome measurements and choose between
two different settings for their measurements. The test is performed by considering many copies of a bipartite
entangled state. The CHSH inequality can be written as
\begin{eqnarray}
 I_{\rm CHSH} & = & p(00|00)+p(00|01)+p(00|10)-p(00|11) \nonumber\\
 & &- p_A(0|0)-p_B(0|0) \stackrel{\mbox{\tiny{ LHV}}}{\leq} 0,
 \label{eq:ICH}
\end{eqnarray}
where $p(ab|xy)$ is the probability that Alice (Bob) obtains the result $a$ ($b$) when performing the measurement $x$ ($y$). $p_A(a|x)$ and $p_B(b|y)$ are the marginal probabilities at Alice and Bob sites, respectively. In quantum mechanics, $p(ab|xy) = {\rm tr} (\rho M_a^x \otimes M_b^y)$, where $\rho$ is a bipartite entangled state, and $M_a^x$ and $M_b^y$ are the measurement operators of Alice and Bob, respectively. The quantum maximum of Eq.~(\ref{eq:ICH}) is $\frac{1}{\sqrt{2}} - \frac{1}{2} \approx 0.207$ \cite{Cirelson80}.

A fundamental point is that inequality (\ref{eq:ICH}) is only valid when the Bell test is performed with perfect detection efficiency, i.e., with $\eta=1$. If all the detectors have the same detection efficiency $\eta$, then the corresponding Bell inequality is \cite{Eberhard}
\begin{eqnarray}
 I_{\rm CHSH} (\eta) &= &\eta^2 I_{\rm CHSH}^{(2)} + \eta(1-\eta)\left(I_{\rm CHSH}^{(1A)} + I_{\rm CHSH}^{(1B)}\right) \nonumber \\
 & &+ (1-\eta)^2I_{\rm CHSH}^{(0)} \stackrel{\mbox{\tiny{ LHV}}}{\leq} 0,
 \label{ICHSHeff}
\end{eqnarray} where $I_{\rm CHSH}^{(2)}$, $I_{\rm CHSH}^{(1A)}$, $I_{\rm CHSH}^{(1B)}$, and $I_{\rm CHSH}^{(0)}$ are, respectively, the values of $I_{\rm CHSH}$, defined in (\ref{eq:ICH}), when two particles, only Alice's particle, only Bob's particle, and no particles are detected. From Eq.~(\ref{ICHSHeff}) one can see that the violation of the CHSH inequality can not be explained in terms of LHV theories only when the overall detection efficiency is above a certain value, $\eta \geq \eta_{\rm req}$, where
\begin{equation}
 \eta_{\rm req} \equiv \frac{p_A(0|0)+p_B(0|0)}{p(00|00)+p(00|01)+p(00|10)-p(00|11)},
 \label{eq:etareq}
\end{equation} which depends on the local measurements chosen. The minimum value of $\eta_{\rm req}$ is usually denoted by $\eta_{\rm crit}$, and it can only be attained with specific measurement settings that require the share of a common reference frame. For any pure bipartite quantum state, there are measurement settings that allows simultaneously for the maximum violation of the CHSH inequality while requiring a detection efficiency $\eta=\eta_{\rm crit}$ \cite{LIVLS12}.

The detection efficiency can be very high in real-world photonic Bell tests, but it is never~1 (e.g., the highest value reported so far is $\eta \approx 0.785$ \cite{GMRWKLCGNUZ12}). Then, the problem is which is the probability of violating the CHSH inequality with random measurements for a given $\eta$.

For solving this problem, we simulate numerically a sufficiently large number of Bell tests with random measurements ($4 \times 10^6$ Bell tests) in each of the three scenarios defined above. Our aim is to obtain numerically with very high precision the probability of violating the CHSH inequality $P_{\rm viol}$ as a function of $\eta$. Similarly to what has been done previously for calculating probabilities of violations with $\eta=1$ \cite{Bartlett10}, our program takes in consideration all equivalent forms of the inequality (\ref{eq:ICH}), i.e., all the different forms one can get for it by relabeling the parties and/or the outcomes and/or the settings. For each interaction, the program generates the random measurements using pseudo-random number subroutines, as in the experiment of Ref.~\cite{Brunner12}, and records the highest value for $I_{\rm CHSH}$ obtained considering all equivalent inequalities. For the set of measurement bases chosen in each run and the inequality with the highest value, the program calculates the required efficiency whenever there is a violation of the CHSH inequality. Each run of the program is independent of the previous interaction. After many runs of the program one obtains a histogram which gives the values of $\eta_{\rm req}$ and the corresponding number of experiments that required such efficiency. Then, from this histogram it is possible to calculate the curve for the $P_{\rm viol}$ versus $\eta$.


\begin{figure}[t]
	\centering
    \includegraphics[width=0.48\textwidth]{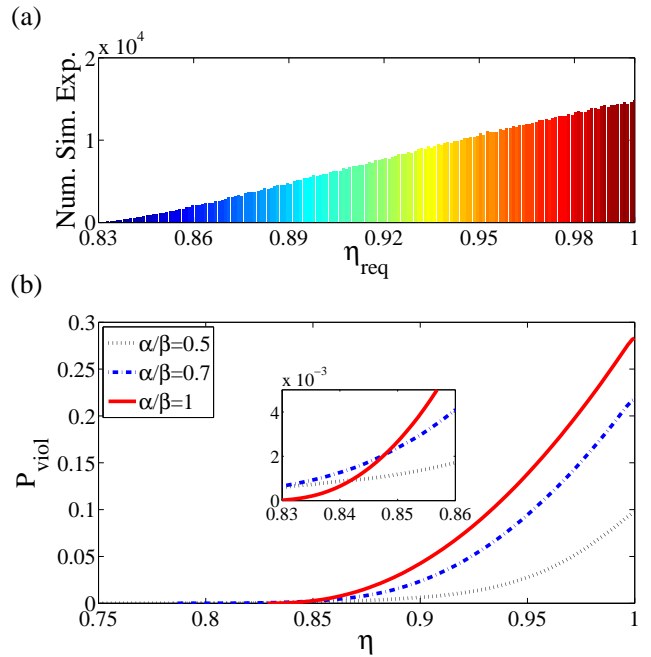}
	\caption{(Color online) (a) Histogram of the required efficiencies (considering the runs with a violation of the CHSH inequality) in the RIM scenario for a MES. The colors in the histograms are related to the numbers of experiments for a certain value of $\eta$. (b) Probability of a loophole-free violation of any of the possible CHSH inequalities as a function of the detection efficiency $\eta$ and for states with different degree of entanglement ($\alpha/\beta$) in Bell tests with RIM. The 3 states represented illustrate the general behavior of $P_{\rm vio}$ vs $\eta$ while increasing or decreasing the entanglement of the initial two-qubit state.}
	\label{RIM}
\end{figure}


We also study the simultaneous behavior of the probability of violation in the three scenarios with $\eta$ and the degree of entanglement of the initial state. For this purpose, we consider two-qubit pure states given by $|\Psi\rangle =\alpha|01\rangle +\beta |10\rangle$, where $\alpha$ and $\beta$ are real and positive. $|0\rangle$ and $|1\rangle$ are the logical states. Therefore, the state with $\alpha/\beta = 1$ corresponds to a maximally entangled state (MES). Those states with values $0<\alpha/\beta<1$ are partially entangled states. This parameter is related with the concurrence $C$ of the state through $\frac{\alpha}{\beta} = \frac{C}{2\beta^2}$ \cite{Wootters2001}.

\section{Probability of a Bell violation with RIM, ROM, and ROTM}

To generate the random measurements corresponding to the RIM case, we use the measurement operators $M_a^x$ and $M_b^y$ as projectors defined according to $|\psi\rangle = \sin\phi|0\rangle + e^{iv_{\phi}}\cos\phi|1\rangle$. Therefore, to have them uniformly distributed over the Bloch sphere, the following condition must hold \cite{Bartlett11}
\begin{equation}
\mbox{RIM} \left\{
\begin{array}{l}
\phi = \frac{1}{2}\arccos(2v - 1) \\
v_{\phi} = 2\pi u
\end{array}
\right.,
\end{equation}
where $u,v \in \{0,1\}$. For one interaction of the program, the values of $u$ and $v$ are randomly chosen for each measurement operator involved in the CHSH test, and the value of $\eta_{\rm req}$ calculated, as mentioned above.

Agreement with previous results has been checked. Specifically, we have checked that our results coincide with those reported in \cite{Bartlett10} in the case of $\eta=1$. In addition, we observed that the expected values of the violations of the CHSH inequality is high enough, such that the violation is robust against noise at the expense of lowering the probability of violation as discussed in \cite{Brunner11}.

Figure~\ref{RIM} summarizes the relevant results for the RIM scenario. In Fig.~\ref{RIM}(a) we show the histogram of $\eta_{\rm req}$ obtained in the simulations for a MES. Figure~\ref{RIM}(b) gives the probability for loophole-free violations of the CHSH inequality as a function of the efficiency for three states with different degree of entanglement. It shows that having a very high detection efficiency is a {\em required resource} for Bell tests in which other resources (namely, device calibration and a common reference frame) are absent, as is the case of the RIM scenario of random Bell tests.

This figure also shows that if we consider the case in which the random Bell test is performed with the best detection efficiency reported so far in a photonic Bell test, i.e., $\eta \approx 0.785$, then there is only a negligible probability of observing a violation that genuinely certifies nonlocality. More specifically, this probability is only $0.005\%$ in the best case, which corresponds to the state with less entanglement defined by $\alpha/\beta = 0.5$. Another example is the following: If we assume a detection efficiency equal to the value of $\eta_{\rm crit}$ required for a conclusive Bell test with MESs, calibrated local measurements and a common reference frame, namely $\eta = 0.828$, then the maximum probability of observing a loophole-free violation in the RIM scenario is only $0.06\%$ [see the box in Fig.~\ref{RIM}(b)].

From Fig.~\ref{RIM}, one reaches the conclusion that if Bell tests of the CHSH inequality are performed without a shared reference frame and without calibrated devices for really certifying nonlocality (and hence secure communications, better-than-classical distributed computation, randomness, or entanglement), then it is necessary to increase substantially the experimental overall detection efficiency (and also the degree of entanglement of the state, with respect to the one which allows for a conclusive Bell test with the lowest detection efficiency $\eta_{\rm crit}=\frac{2}{3}$ \cite{Eberhard}). Specifically, in the RIM case, for having a probability of violation higher than $5\%$ we need a detection efficiency of at least $\eta = 0.90$ for the MES.


\begin{figure}[t]
	\centering
    \includegraphics[width=0.48\textwidth]{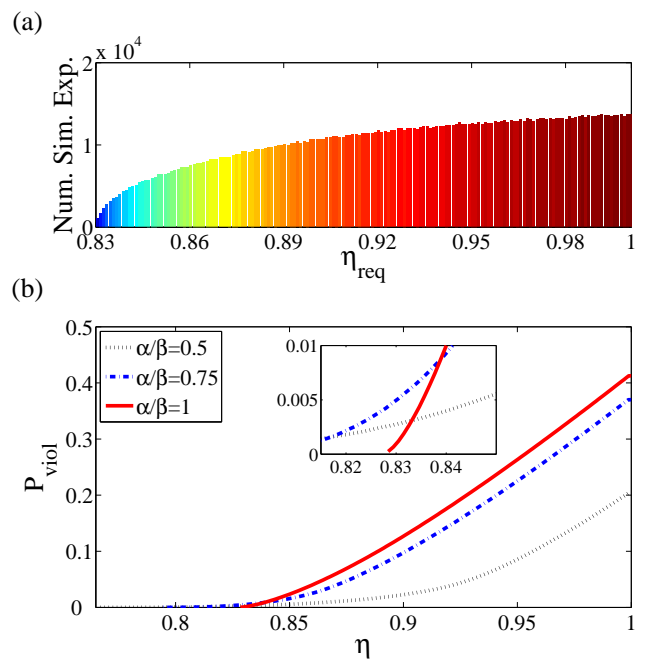}
	\caption{(Color online) (a) Histogram of the required efficiencies (considering the runs with a violation of the CHSH inequality) in the ROM scenario for a MES. The colors in the histograms are related to the numbers of experiments for a certain value of $\eta$. (b) Probability of a loophole-free violation of any of the possible CHSH inequalities as a function of the detection efficiency $\eta$ and for states with different degree of entanglement ($\alpha/\beta$) in Bell tests with ROM. The 3 states represented illustrate the general behavior of $P_{\rm vio}$ vs $\eta$ while increasing or decreasing the entanglement of the initial two-qubit state.}
	\label{ROM}
\end{figure}


Now let us consider the results obtained within the ROM scenario. In Fig.~\ref{ROM}(a) we show the histogram of $\eta_{\rm req}$ obtained in the simulations for a MES. This histogram has a different behavior than the one of the RIM scenario [Fig.~\ref{RIM}(a)]. For example, one can see that a higher number of experiments demanded lower required efficiencies. The resulting probability for loophole-free violations of the CHSH inequality as a function of $\eta$ and for states with different degree of entanglement is shown in Fig.~\ref{ROM}(b). To obtain these results we used the program described above, but now with the constraint that the measurement directions associated to the operators $M_{a=0}^{x=0}$ ($M_{b=0}^{y=0}$) and $M_{a=0}^{x=1}$ ($M_{b=0}^{y=1}$) are orthogonal. This restriction is implemented by taking the first operator's eigenstate randomly and uniformly distributed over the Bloch sphere and forcing the second operator's eigenstate to lie in an orthogonal plane. The second eingestate direction is also randomly taken in this orthogonal plane.

Figure~\ref{ROM}(b) shows that, for a given $\eta$, the ROM scenario allows for a higher probability of violating the CHSH inequality. The reason explaining this is that MESs are the most robust states against randomization of the local measurements and that for these states the maximal violation of the CHSH inequality is obtained when orthogonal measurements are used. Notice, however, that unlike the RIM scenario, the ROM scenario requires the extra resource of using calibrated devices. Even though the probability for a genuine CHSH-violation is higher in the ROM scenario than in the RIM case, it is still very low even for very high values of $\eta$. For example, when $\eta=0.785$ and $\eta=0.828$, the highest probability is only $0.015\%$ and $0.042\%$, respectively.


\begin{figure}[t]
	\centering
    \includegraphics[width=0.48\textwidth]{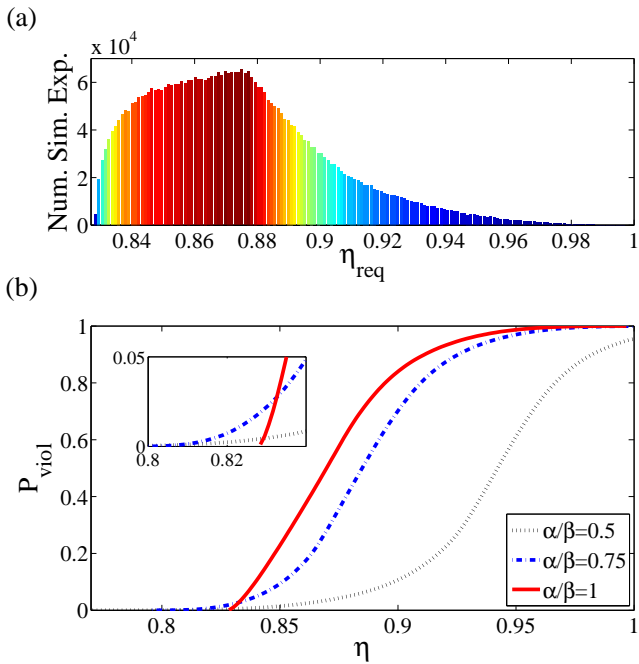}
	\caption{(Color online) (a) Histogram of the required efficiencies in the ROTM scenario for a MES. The colors in the histograms are related to the numbers of experiments for a certain value of $\eta$. (b) Probability of a loophole-free violation of any of the possible CHSH inequalities as a function of the detection efficiency $\eta$ and for states with different degree of entanglement ($\alpha/\beta$) in Bell tests with ROTM. The 3 states represented illustrate the general behavior of $P_{\rm vio}$ vs $\eta$ while increasing or decreasing the entanglement of the initial two-qubit state.}
	\label{ROTM}
\end{figure}


For the ROTM scenario we repeated our investigation and the resulting histogram obtained for the required efficiencies for the MES is given in Fig.~\ref{ROTM}(a). It has the shape of the curve obtained in \cite{Brunner11} while studying the CHSH-violations values in this scenario for a MES. The curve of the probability for a loophole-free CHSH-violation as a function of $\eta$, and for states with different degree of entanglement is shown in Fig.~\ref{ROTM}(b). One can see that the behavior is similar to the ones of Figs.~\ref{RIM} and \ref{ROM}, showing that even though this scenario allows for higher violating probabilities, it also demands the use of higher detection efficiency. For example, for $\eta=0.828$, the probability of violating the inequality is only $1.8\%$.


\section{Conclusions}

Progress in quantum information processing requires the identification of which resources are actually critical for real-world applications. Bell tests are powerful tools for certifying nonlocality, communication security, better-than-classical distributed computation, randomness, and entanglement. Here we have shown that the conclusions reached in recent investigations demonstrating that neither perfect device calibration nor common reference frames are essential for successful Bell tests have overlooked the role of the detection efficiency.

For all previous introduced scenarios of random Bell tests \cite{Bartlett10,Bartlett12,Brunner11} we have obtained the dependence of the violating probabilities of the CHSH-inequality with the overall detection efficiency, while considering states with different degree of entanglement. As a side-project of our work we observed that only when the detection efficiencies are very high, that the violating probabilities become relevant (independently of the scenario considered). This emphasizes the importance of frame synchronization, since the required efficiencies of random Bell tests are still out of what is experimentally possible even with the state-of-the-art of photodetectors.

While our investigation was focused on three types of Bell tests with random
measurements, our results can be extrapolated to more common scenarios such as the one in which
there are random drifts on the measurement parameters over the experimental process. For example, the RIM scenario we discussed can be seen as an extreme case in which there are always drifts on the measurement parameters. From the results obtained, one may expect that, if the experiment is susceptible
to random drifts, then higher efficiencies will be required for a loophole-free Bell violation.


\begin{acknowledgments}
This work was supported by Grants FONDECYT 1120067, Milenio~P10-030-F, PIA-CONICYT~PFB0824 (Chile), the Science without Borders Program (Capes and CNPq, Brazil), and the Project No.\ FIS2011-29400 (MINECO, Spain). E.S.G., G.C. and J.F.B. acknowledge the financial support of CONICYT and AGCI. A.C. acknowledges the Universidad de Concepci\'{o}n for its hospitality.
\end{acknowledgments}



\begin{thebibliography}{99}

\bibitem{Bell}
 J. S. Bell, Physics \textbf{1}, 195 (1964).


\bibitem{Ekert91}
 A. K. Ekert,
 Phys. Rev. Lett. \textbf{67}, 661 (1991).

\bibitem{BCMD10}
 H. Buhrman, R. Cleve, S. Massar, and R. de Wolf,
 Rev. Mod. Phys. \textbf{82}, 665 (2010).

\bibitem{Acin}
 A. Ac\'{i}n, S. Massar, and S. Pironio,
 Phys. Rev. Lett. \textbf{108}, 100402 (2012).

\bibitem{GT09}
 O. G\"uhne and G. T\'oth,
 Phys. Rep. \textbf{474}, 1 (2009).


\bibitem{Lima2010}
 G. Lima, G. Vallone, A. Chiuri, A. Cabello, and P. Mataloni,
 Phys. Rev. A \textbf{81}, 040101(R) (2010).

\bibitem{Chaves11}
 R. Chaves and J. B. Brask,
 Phys. Rev. A \textbf{84}, 062110 (2011).

\bibitem{Smith2012}
 D. H. Smith, G. Gillett, M. P. de Almeida, C. Branciard, A. Fedrizzi,
 T. J. Weinhold, A. Lita, B. Calkins, T. Gerrits, H. M. Wiseman, S. W. Nam, and A. G. White,
 Nature Comm. \textbf{3}, 625 (2012).

\bibitem{Zeilinger12}
 B. Wittmann, S. Ramelow, F. Steinlechner,
 N. K Langford, N. Brunner, H. M. Wiseman, R. Ursin, and A. Zeilinger,
 New J. Phys. \textbf{14}, 053030 (2012).

\bibitem{CS12}
 A. Cabello and F. Sciarrino,
 Phys. Rev. X \textbf{2}, 021010 (2012).

\bibitem{GMRWKLCGNUZ12}
 M. Giustina, A. Mech, S. Ramelow, B. Wittmann,
 J. Kofler, A. Lita, B. Calkins, T. Gerrits,
 S. W. Nam, R. Ursin, and A. Zeilinger,
 Nature \textbf{497}, 227-230 (2013).

\bibitem{Aspect} A. Aspect, J. Dalibard, and G. Roger, Phys. Rev. Lett. \textbf{49}, 1804-1807 (1982).

\bibitem{Weihs} G. Weihs \textit{et al}., Phys. Rev. Lett. \textbf{81}, 5039 (1998).


\bibitem{Pearle}
 P. M. Pearle,
 Phys. Rev. D \textbf{2}, 1418 (1970).


\bibitem{Adan03}
 A. Cabello,
 Phys. Rev. Lett. \textbf{91}, 230403 (2003).

\bibitem{Cabello03}
 A. Cabello,
 Phys. Rev. A \textbf{68}, 042104 (2003).


\bibitem{Bartlett10}
 Y.-C. Liang, N. Harrigan, S. D. Bartlett, and T. Rudolph,
 Phys. Rev. Lett. \textbf{104}, 050401 (2010).

\bibitem{Bartlett11}
 J. J. Wallman, Y.-C. Liang, and S. D. Bartlett,
 Phys. Rev. A \textbf{83}, 022110 (2011).


\bibitem{Bartlett12}
 J. J. Wallman and S. D. Bartlett,
 Phys. Rev. A \textbf{85}, 024101 (2012).


\bibitem{Brunner11}
 P. Shadbolt, T. V\'{e}rtesi, Y.-C. Liang, C. Branciard, N. Brunner, and J. O'Brien,
 Scientific Reports \textbf{2}, 470 (2012).

 \bibitem{Brunner12}
 J. B. Brask, R. Chaves, and N. Brunner,
 \eprint{arXiv:12122019}.

\bibitem{Wallman}
 M. S. Palsson, J. J. Wallman, A. J. Bennet, and G. J. Pryde,
 Phys. Rev. A \textbf{86}, 032322 (2012).


\bibitem{CHSH}
 J. F. Clauser, M. A. Horne, A. Shimony, and R. A. Holt,
 Phys. Rev. Lett. \textbf{23}, 880 (1969).

\bibitem{ClauserHorne}
 J. F. Clauser and M. A. Horne,
 Phys. Rev. D \textbf{10}, 526 (1974).


\bibitem{BrunnerPLA}
 N. Brunner and N. Gisin,
 Phys. Lett. A \textbf{372}, 3162 (2008).

\bibitem{CRV08}
 A. Cabello, D. Rodr\'{\i}guez, and I. Villanueva,
 Phys. Rev. Lett. \textbf{101}, 120402 (2008).

\bibitem{Vertesi2009}
 K. P\'{a}l and T. V\'{e}rtesi,
 Phys. Rev. A \textbf{79}, 022120 (2009).

\bibitem{CLR09}
 A. Cabello, J.-\AA. Larsson, and D. Rodr\'{\i}guez,
 Phys. Rev. A \textbf{79}, 062109 (2009).


\bibitem{Cirelson80}
 B. S. Cirel'son,
 Lett. Math. Phys. \textbf{4}, 93 (1980).


\bibitem{Eberhard}
 P. H. Eberhard,
 Phys. Rev. A \textbf{47}, R747 (1993).


\bibitem{LIVLS12}
 G. Lima, E. B. Inostroza, R. O. Vianna, J.-\AA. Larsson, and C. Saavedra,
 Phys. Rev. A \textbf{85}, 012105 (2012).


\bibitem{Wootters2001}
 W. K. Wootters,
 Quant. Inf. Comp. \textbf{1}, 27 (2001).

\end{thebibliography}
\end{document}